\documentclass{article}
\pdfoutput=1
\usepackage{graphicx}
\usepackage{subcaption}
\usepackage[includefoot,margin=1in]{geometry} 

\usepackage{tikz}
\usetikzlibrary{arrows.meta}
\usetikzlibrary{plotmarks}
\usetikzlibrary{shapes}
\usepackage{pgfplots}
\pgfplotsset{compat=newest}
 
\usepackage{array} 
\usepackage{footnote}
 
\usepackage{amssymb}
\usepackage{amsmath}
\usepackage{amsfonts}
\usepackage{bm}

\usepackage{url}
\usepackage{hyperref}
\hypersetup{pdfstartview={XYZ null null 1.00}}

\newcommand{\beq}{\begin{equation}}
\newcommand{\eeq}{\end{equation}}

\newcommand{\beqr}{\begin{eqnarray}}
\newcommand{\eeqr}{\end{eqnarray}}
\def\bal#1\eal{\begin{align}#1\end{align}}
\def\bat#1#2\eat{\begin{alignat}{#1}#2\end{alignat}}

\begin{document}
\setlength{\emergencystretch}{3em}

\title{An origami Universal Turing Machine design}
\author{Michael Assis\footnote{Department of Obstetrics, Gynaecology and Newborn Health, University of Melbourne, Parkville, VIC, Australia}
}
\date{\today}
\maketitle

\begin{abstract}
It has been known since 1996 that deciding whether a collection of creases on a piece of paper can be fully folded flat without causing self-intersection or adding new creases is an NP-Hard problem (Bern and Hayes). In their proof, a binary state was implemented as a pleat, with the state corresponding to the pleat layering order; states then interact via pleat intersections. Building on some of the machinery of their result, we will present a method for constructing an origami NAND logic gate, leading to a theoretical origami Universal Turing Machine.
\end{abstract}

\section{Introduction}
Origami has been known to be computationally complex since the work of Bern and Hayes, where they showed that deciding whether folding a given collection of creases will result in a flat-foldable state is an NP-Hard problem~\cite{bern1996,hull2011}. Naturally it could be asked whether origami possesses enough complexity to be able to perform arbitrarily complex calculations, that is, whether a Universal Turing Machine could be constructed from paper using origami. 

Turing machines are theoretical models of computation. They are simple machines that manipulate symbols on an infinite strip of tape according to rules chosen for the machine. Depending on an initial configuration of symbols on the tape, the machine will produce different printed outcomes. While there could in principle be an infinite number of distinct Turing machines that could produce unique and incompatible results, Universal Turing Machines~(UTMs) exist that allow one to simulate any other Turing machine. Therefore, showing that a Universal Turing Machine can be implemented, such as in origami, is equivalent to showing that origami can compute any computable function, according to the Church-Turing Thesis~\cite{turing1937}.

There are many equivalent methods for implementing a UTM. One method is to implement a Turing-complete language. While modern programmers use languages like C++, Python, or Java, we would want the simplest possible language to implement. Among the simplest, it would be useful to require that they not have Go To statements, to avoid jumps to arbitrary locations, and that they operate on binary symbols. One such language is P’’~\cite{bohm1964,bohm1966}, which has rules for moving left/right one cell on the tape, swapping the 0/1 symbols at the current cell, and has while loops. The implementation of while loops is probably the most non-trivial aspect of P’’, since it requires some kind of logic to decide when to break the loop. And as long as logic needs to be implemented, we can shift the focus to implementing logic gates, such as NAND gates. Our method of constructing a UTM, then, is to construct an origami NAND gate and memory rather than implementing a Turing-complete language. 

We note that recently a similar result was published in a pre-print~\cite{hull2023}, where they implement the Turing-complete Rule 101~\cite{cook2004} in origami. The present work was completed and presented at two conferences in 2021~\cite{Assis2021,Assis2021b}, and while video footage was recorded, this is the first printed publication of the result, with our method also differing from that of~\cite{hull2023}.

We begin by introducing our method of constructing a NAND gate, followed by the implementation of memory, and we follow with a discussion of the efficiency of our approach, a discussion of rigid foldability, and we compare our method to that of~\cite{hull2023}. 

\section{Implementing a NAND gate in origami}
Rather than implementing all types of logic gates, it is sufficient to implement a NAND gate, which is functionally complete, that is, all other logic gates can be constructed by cleverly connecting a series of NAND gates~\cite{teuscher2002}. 

Borrowing some of the methods from Bern and Hayes, we choose to define our binary states 0 and 1 in terms of the layer ordering of a pleat, parallel mountain and valley folds, together with a chosen direction, as shown in Figure~\ref{fig:states} below. 
\begin{figure}[htpb]
\begin{center}
\includegraphics[height=0.6in]{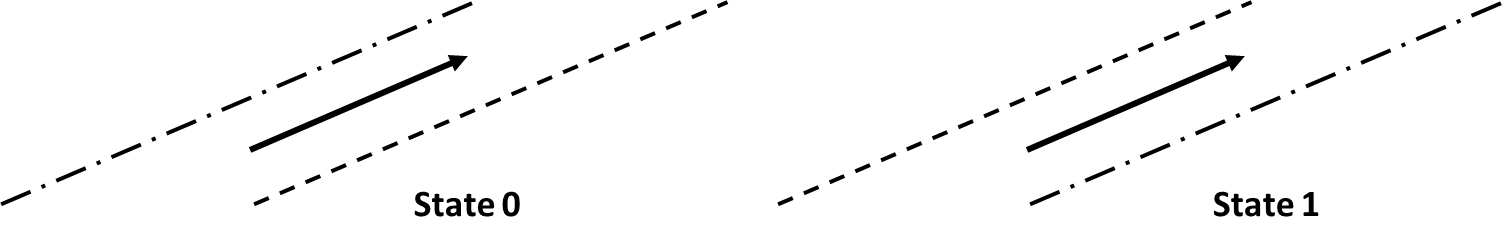}
\end{center}
\caption{Definition of the binary states 0 and 1 on the left and right, respectively.\label{fig:states}}
\end{figure}

\begin{figure}[htpb]
\begin{center}
\includegraphics[height=1.5in]{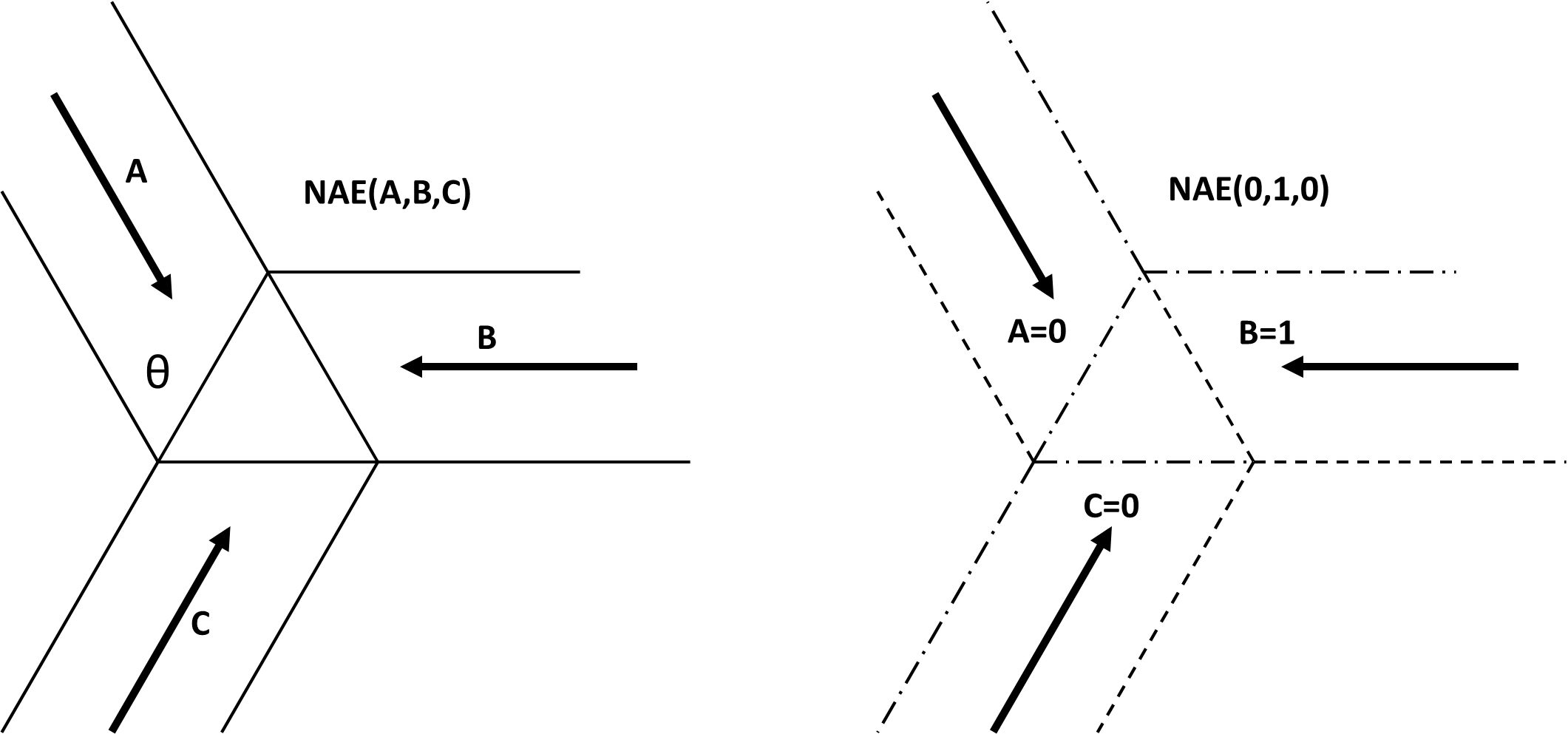}
\end{center}
\caption{The NAE clause of Bern and Hayes with three incoming states A, B, C, defined on the left. The clause is foldable when the signals A, B, C are not all equal and for $\theta>30^{\circ}$. On the right is the clause NAE(0,1,0) that can be folded flat.\label{fig:NAE}}
\end{figure}

In their work, Bern and Hayes implemented a not-all-equal (NAE) clause gadget that accepts three input signals, shown in Figure~\ref{fig:NAE}. The NAE clause can only be folded flat when the three incoming states are not all equal to each other and by choosing an angle $\theta>{30}^{\circ}$. The angle between the pleats is always $120^{\circ}$ regardless of the angle $\theta$ chosen, and if the angle $\theta=60^{\circ}$ is chosen for the gadget, then the pleats can lie along a triangular grid as is common in origami tessellations.

\begin{figure}[htpb]
\begin{center}
\includegraphics[height=1.5in]{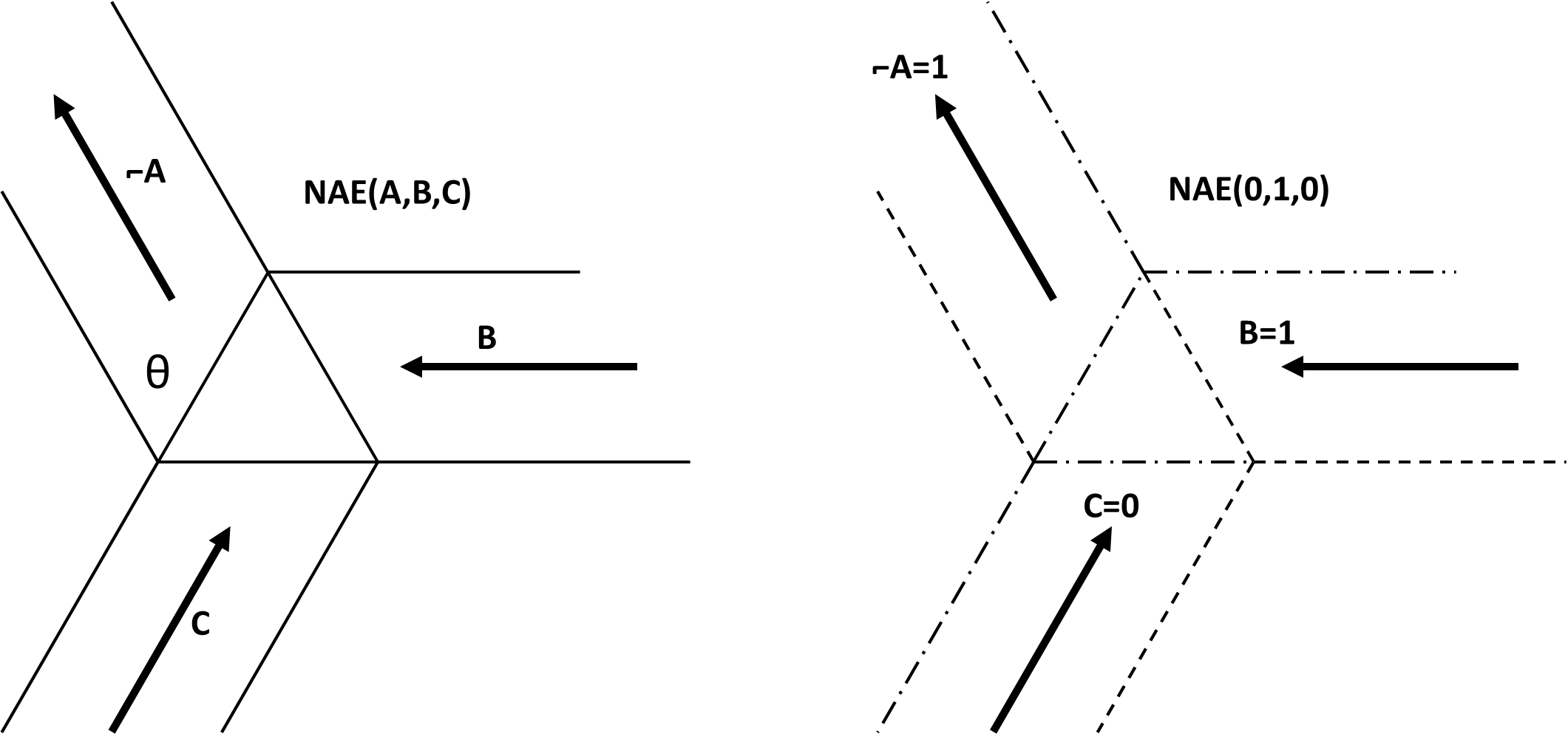}
\end{center}
\caption{The NAE clause can be re-interpreted as two incoming states and one outgoing negated state, shown on the left. On the right the clause NAE(0,1,0) with the outgoing negated state, $\neg$A=1.\label{fig:NAEredef}}
\end{figure}

We choose to re-interpret the NAE clause, however, to have two incoming signals and one outgoing signal, where the outgoing signal is negated compared to the signal used in the NAE clause, as shown in Figure~\ref{fig:NAEredef} where the NAE(A,B,C) clause has outgoing signal $\neg$A. This is useful because we don’t wish to have signals terminate, but rather that the NAE will be used in the construction of the NAND gate.

\begin{figure}[htpb]
\begin{center}
\includegraphics[height=1.5in]{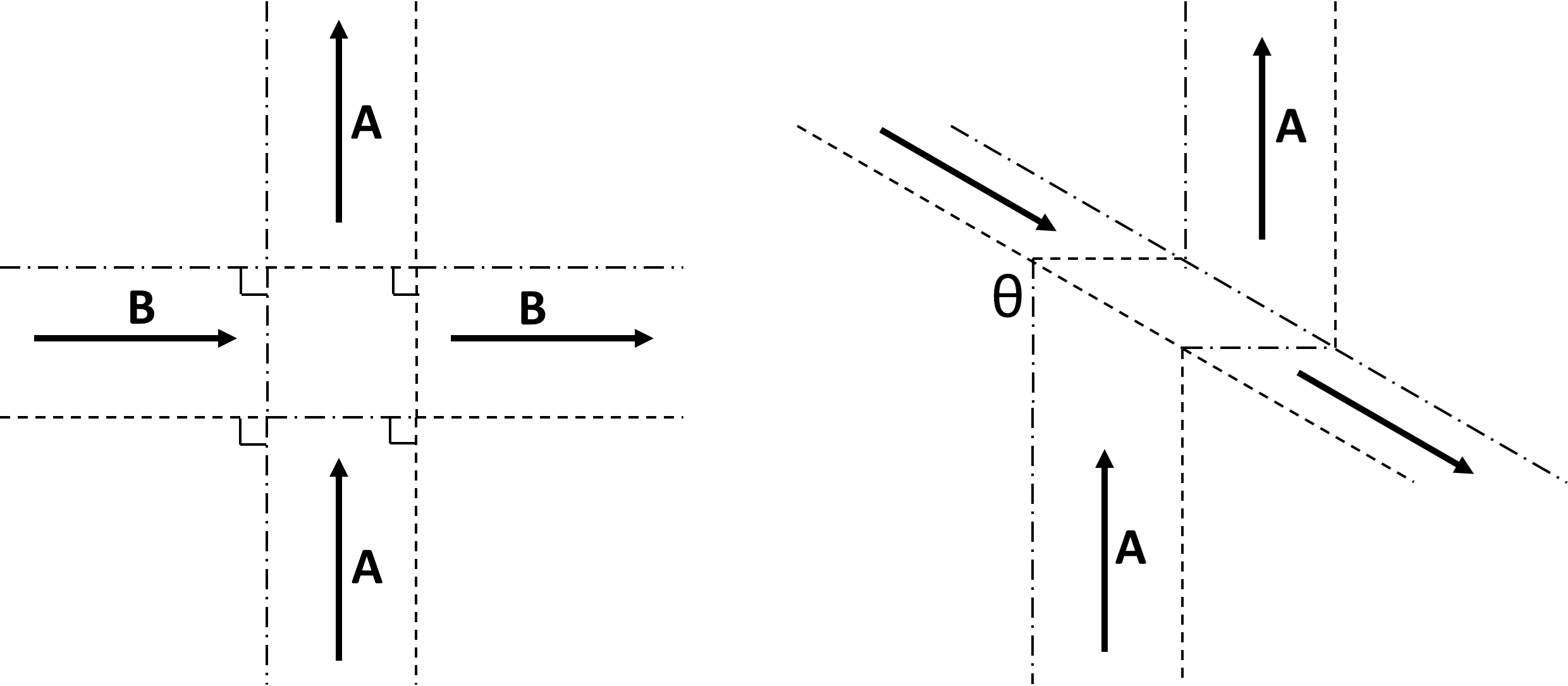}
\end{center}
\caption{Two signals A and B can pass through each other unhindered when intersecting at right angles (left), otherwise, only the pleat that zig-zags, A, can pass through unhindered (right).\label{fig:crossings}}
\end{figure}

As more and more signals fill the paper, one might worry about the interactions of signals that cross each other. As found in~\cite{akitaya2016}, if pleats intersect at a right angle, the pleats can pass through each other unimpeded with their states unchanged. Otherwise, as seen in Figure~\ref{fig:crossings}, only the pleat that zig-zags can pass through unimpeded, the other pleat possibly changing its state. Therefore, for any two pleats for which we are tracking their states, they must intersect at right angles. If a signal we are tracking is crossing a pleat we are disregarding, then the pleats can intersect at other angles so long as the pleat we are tracking is the one to zig-zag through the other pleat. And either choice can be made in the case of two pleats whose states we are not interested in.

\begin{figure}[htpb]
\begin{center}
\includegraphics[height=1.5in]{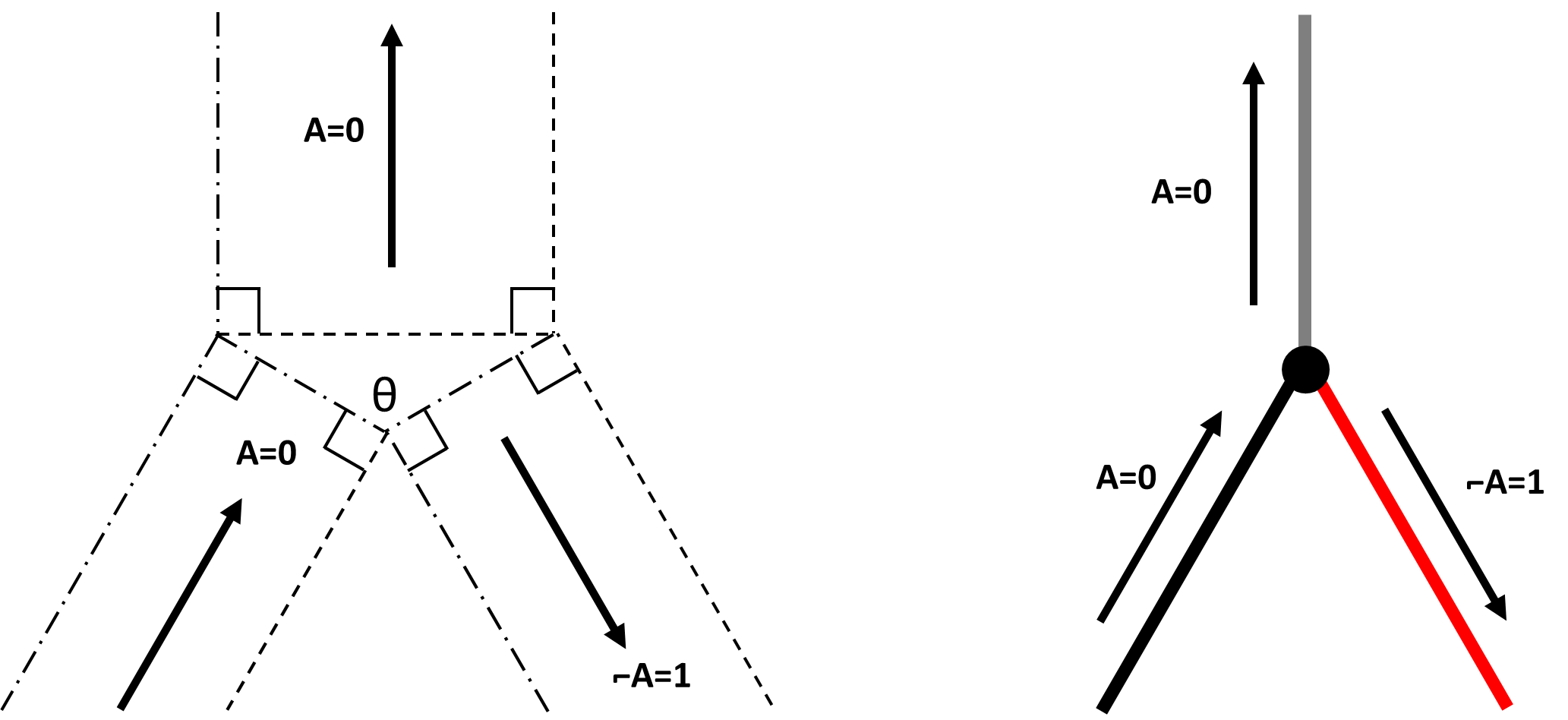}
\end{center}
\caption{The reflector gadget of Bern and Hayes on the left, with one input signal and two output signals, one equal to the input but with wider pleat width and one negated of equal pleat width. It folds flat for $\theta>90^{\circ}$. On the right our drawing convention, with black for signals of interest, gray for intermediate signals, red for signals to be ignored, and the reflector gadget as a filled circle. \label{fig:reflector}}
\end{figure}

The final gadget we wish to borrow from Bern and Hayes is their reflector gadget, as shown in Figure~\ref{fig:reflector}. The reflector accepts an incoming signal and outputs two signals, one equal to the input but with a wider pleat width, and one negated and with the same pleat width. The reflector gadget folds flat as long as the angle $\theta>90^{\circ}$, and we could choose $\theta=120^{\circ}$ for convenient placement of the signals along a triangular grid. In Figure~\ref{fig:crossings} we also define a drawing convention where we collapse the pleats to a single line and represent the reflector gadget by a filled circle. We use the color convention that black represents the signal of interest, gray represents intermediate transmission steps of the signal, and red represents extraneous signals which can be ignored and will go out to the border of the paper. 

\begin{figure}[htpb]
\begin{center}
\includegraphics[height=2in]{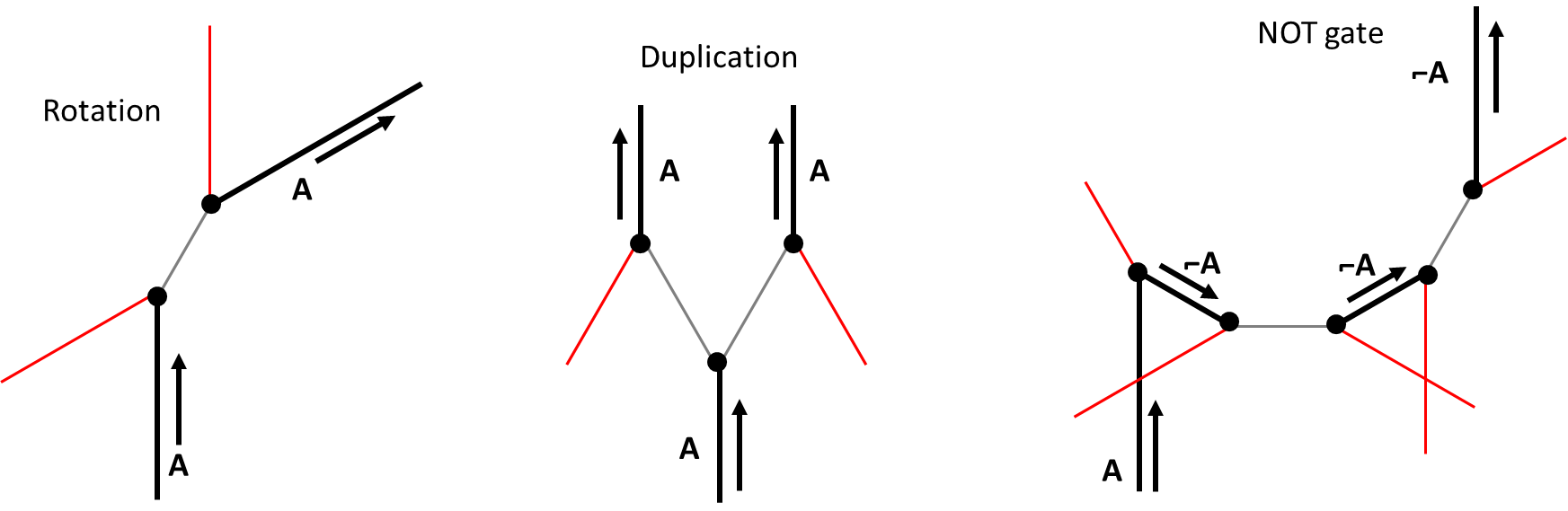}
\end{center}
\caption{On the left a signal rotator gadget, in the middle a gadget to duplicate a signal, and on the right a NOT gate gadget. The duplicator gadget in reverse is a combiner of equal signals. In all cases the incoming and outgoing signals remain the same width.  \label{fig:newgadgets}}
\end{figure}

We now define three new gadgets using the reflector gadget: a rotation gadget, a duplication gadget and a NOT gate gadget, as shown in Figure~\ref{fig:newgadgets}. Note that while the reflector gadget itself is able to create negations, our NOT gadget allows the signal to continue in the same initial direction. Note also that the duplication gadget can be viewed in reverse as a combiner of two equal signals. All of our gadgets also have the property that the incoming and outgoing pleat widths are the same size. 

We are now in the position to construct our NAND gate, which has the truth table shown in Table~\ref{tab:nand}.
\begin{table}
\begin{center}
\caption{The NAND gate truth table.\label{tab:nand}}
\begin{tabular}{ c | c | c }
  A & B & A NAND B \\\hline
  0 & 0 & 1 \\
  0 & 1 & 1 \\
  1 & 0 & 1 \\
  1 & 1 & 0 
\end{tabular}
\end{center}
\end{table}

We would like to use NAE gadgets to construct the NAND truth table for two incoming signals. We can accomplish this by using 3 NAE gadgets and three incoming signals A, B, and an auxiliary signal P. This is the minimum number of NAE gadgets and signals that can construct a NAND gate. To see this, we first list the full truth table for 3 input signals and one output signal in Table~\ref{tab:three}.
\begin{table}
\begin{center}
\caption{The truth table for three input signals A, B, P, and one output signal O. The columns in bold give O=A NAND B.\label{tab:three}}
\begin{tabular}{ |c|c|c|c|c|c|c|c|c|c|c|c|c|c|c|c|c| }
\hline
A & 1 & 1 & 1 & \textbf{1} & 1 & 1 & \textbf{1} & 1 & 0 & 0 & \textbf{0} & 0 & 0 & 0 & \textbf{0} & 0 \\\hline
B & 1 & 1 & 1 & \textbf{1} & 0 & 0 & \textbf{0} & 0 & 1 & 1 & \textbf{1} & 1 & 0 & 0 & \textbf{0} & 0 \\\hline
P & 1 & 1 & 0 & \textbf{0} & 1 & 1 & \textbf{0} & 0 & 1 & 1 & \textbf{0} & 0 & 1 & 1 & \textbf{0} & 0 \\\hline
O & 1 & 0 & 1 & \textbf{0} & 1 & 0 & \textbf{1} & 0 & 1 & 0 & \textbf{1} & 0 & 1 & 0 & \textbf{1} & 0 \\\hline
\end{tabular}
\end{center}
\end{table}

We would like to choose NAE gadgets with combinations of the three input signals A, B, P such that the only truth table columns which are valid are those which result in an output signal O equal to A NAND B. We can accomplish this by having the following three NAE gadgets being simultaneously satisfied and also requiring that the signal P always equals 0:
\beq
\mathrm{NAE(A,B,O)},\quad \mathrm{NAE(A,P,O)},\quad \mathrm{NAE(B,P,O)}
\eeq
With these choices, all columns in the truth table are eliminated except the bold columns for which the columns A, B, O give the NAND truth table. In other words, the combination of these three gadgets and an auxiliary signal P that is always 0 will only fold flat if the output follows the NAND gate truth table for input signals A and B.
 
In order for this to work, the same signals A, B, P need to appear in multiple NAE gates, which requires duplication. And in terms of geometry, we will need to rotate signals so that they can reach the relevant NAE gate. A schematic to accomplish this is sketched in Figure~\ref{fig:nandschematic} below. 
\begin{figure}[htpb]
\begin{center}
\includegraphics[height=3in]{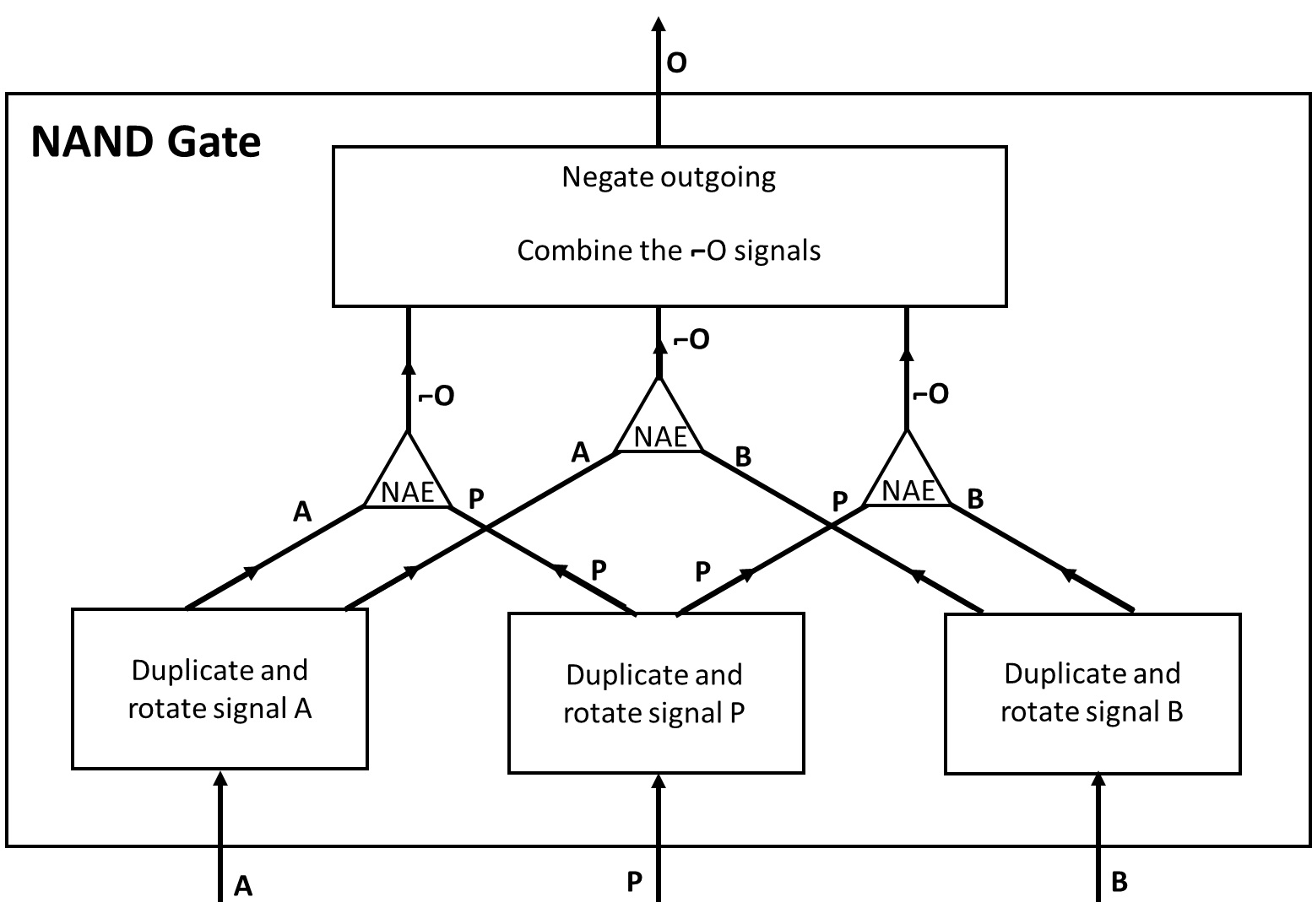}
\end{center}
\caption{A schematic diagram showing how the input signals A, B, together with the auxiliary signal P, are connected inside the NAND gate to produce output signal O = A NAND B. \label{fig:nandschematic}}
\end{figure}

\begin{figure}[htpb]
\begin{center}
\includegraphics[height=4in]{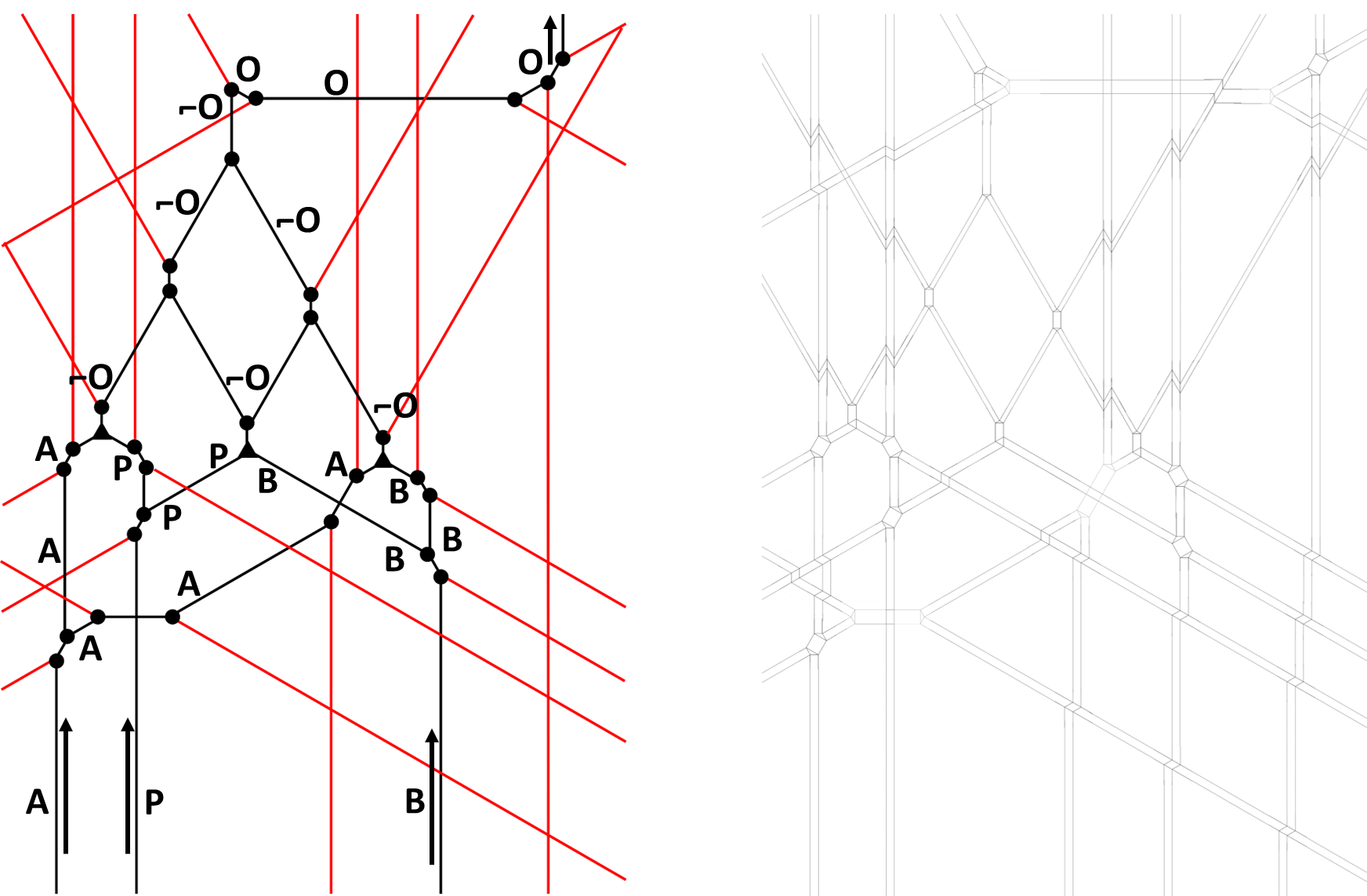}
\end{center}
\caption{The crease pattern schematic (left) and full crease pattern (right) of the origami NAND gate.  \label{fig:nandfull}}
\end{figure}

The signal P is a necessary signal in our NAND gate that is nevertheless always chosen to be 0. One single such signal can be propagated and duplicated throughout the entire machine, making sure to never flip the initial signal. In this sense, it is similar to a constant voltage signal that powers a computer, which ceases to work once the power switch is turned off. 

Figure~\ref{fig:nandfull} shows a schematic and a full crease pattern for the NAND gate. The signals of interest are in black, with red representing states we’re not interested in. The triangles represent the NAE gadgets, and dots represent reflector gadgets forming rotation, duplication, and NOT gadgets.

\section{From a NAND gate to a Universal Turing Machine}
\begin{figure}[htpb]
\begin{center}
\includegraphics[height=1in]{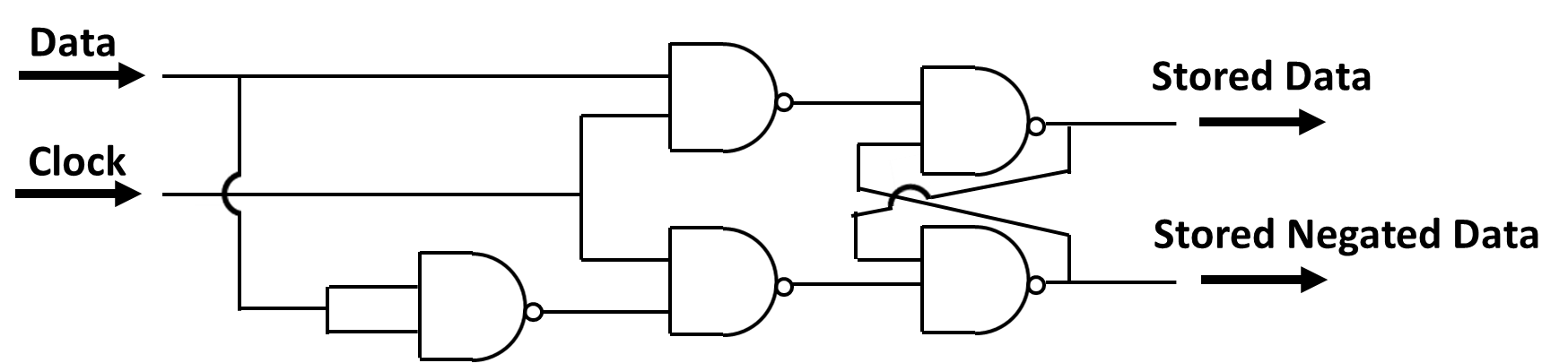}
\end{center}
\caption{The crease pattern schematic on the left and full crease pattern on the right of the origami NAND gate.  \label{fig:dflipflop}}
\end{figure}

Now that we have an origami NAND gate, we can duplicate it across the paper as many times as we wish in order to generate all possible logic gates for building a CPU. However, in order to build a UTM we also need to have memory. Memory can be built from D-type flip-flop registers, which themselves can be built from a NAND gate together with a clock signal, as shown in Figure~\ref{fig:dflipflop}. The clock signal, therefore, is the last component we will need. In our machine, one such signal can be duplicated to service the entire computer, with its signal pleat flipping its state at regular intervals. The interval can be chosen long enough so that all relevant intermediate folds have been processed and the computer is ready to receive the next clock cycle. 

Therefore, as long as sufficient NAND gates are folded and connected appropriately with the P and clock signals, memory registers can be built, CPU instructions can be defined, a bus can be implemented, a multiplexer constructed, etc~\cite{teuscher2002}. In other words, all the elements of a modern computer. 

More theoretically, as long as we have infinite sheet of paper with an infinite number of NAND gates connected properly with an infinite amount of memory, we will have constructed a Universal Turing Machine.

\section{Discussion}
In terms of efficiency of our NAND gate, there are 2 input signals, an auxiliary signal, and an output signal, plus 22 extraneous pleats, for an efficiency of 18\% in terms of number of pleats, though efficiency of folding scales with pleat length.  These extraneous signals can be propagated outward to the paper border avoiding relevant parts of the machine, but they will still need to be folded. In that sense, they are like heat generated in conventional computers, which is created during computation but which serves no purpose of itself. It would be interesting to determine the theoretical limit of efficiency of an origami-based computer. 

It could be also asked whether a rigidly foldable origami computer can be constructed. Whether or not it can be, the current gadgets presented here don’t allow it. This is because no rigidly foldable triangle twist exists~\cite{lang2018}, which is required for our NAE gadgets. One can construct a NAE 4 clause rigidly foldable gadget, but it’s unclear whether rigidly foldable reflectors could be found. Furthermore, we wouldn’t want the entire paper to unfold each time one small part is folded. Therefore, rigidly foldable structures would need to be found which have more than one degree of freedom so that parts of the computer could be isolated from other parts in their folding. This requires the use of gadgets with vertices having more than 4 creases intersecting them~\cite{lang2018}. Recently~\cite{treml2018} considered individual rigidly foldable logic gates made from box-pleating crease patterns with vertices of degree 8. More work would need to be done to see whether a full rigidly foldable computer is possible. 

Concerning the recent method of showing Turing Completeness by~\cite{hull2023}, we note that their work produces a much smaller NAND gate than ours, being a single gadget. Conceptually their method is simple, requiring the same Rule 101 implemented in all cells of the paper, although our duplicator and rotator gadgets allow a greater flexibility in design, allowing for the building a computer directly analogous to conventional computers. While both methods of proving that origami can compute arbitrarily complex computations is of theoretical interest, both will likely remain as impracticable methods for performing even very elementary computations. 

\section*{Acknowledgements}
The author would like to thank the anonymous referee and Thomas Hull for their suggestions to improve the paper, in particular pointing out the reflector gadget issue originally discovered by Akitaya et al in 2016~\cite{akitaya2016}, with the result being that the NAND gate received two extra rotation gadgets and a simpler to fold crease pattern was chosen. 


\end{document}